\documentclass[twocolumn,showpacs,preprintnumbers]{revtex4}
\usepackage{amssymb}
\usepackage{amsmath}
\usepackage{dcolumn}
\usepackage{bm}
\usepackage{graphicx}
\usepackage[dvipdfm,colorlinks=true, citecolor=blue, urlcolor=blue ]{hyperref}

\begin{document}

\title{Preservation of quantum correlation between separated
nitrogen-vacancy centers embedded in photonic crystal cavities}
\author{W. L. Yang$^{1,3}$}
\author{Jun-Hong An$^{2,3}$}
\email{anjhong@lzu.edu.cn}
\author{Chengjie Zhang$^{3}$}
\author{M. Feng$^{1}$}
\email{mangfeng@wipm.ac.cn}
\author{C. H. Oh$^{3}$}
\email{phyohch@nus.edu.sg}
\affiliation{$^{1}$ State Key Laboratory of Magnetic Resonance and Atomic and Molecular
Physics, Wuhan Institute of Physics and Mathematics, Chinese Academy of
Sciences, Wuhan 430071, China}
\affiliation{$^{2}$ Center for Interdisciplinary Studies, Lanzhou University, Lanzhou
730000, China}
\affiliation{$^{3}$ Centre for Quantum Technologies and Department of Physics, National
University of Singapore, Singapore 117543, Singapore}

\begin{abstract}
We investigate the non-Markovian dynamics of quantum correlation between two
initially entangled nitrogen-vacancy centers (NVC) embedded in photonic
crystal cavities (PCC). We find that a finite quantum correlation is
preserved even asymptotically when the transition frequency of the NVC is
within the band gap of the PCC, which is quantitatively different from the
result of approaching zero under the Born-Markovian approximation. In
addition, once the transition frequency of NVC is far beyond the bad gap of
the PCC, the quantum correlation initially prepared in NVC will be fully
transferred to the reservoirs in the long-time limit. Our result reveals
that the interplay between the non-Markovian effect of the structured
reservoirs and the existence of emitter-field bound state plays an essential
role in such quantum correlation preservation. This feature may open new
perspectives for devising active decoherence-immune solid-state optical
devices.
\end{abstract}

\maketitle

\section{Introduction}
Recently, decoherence control and non-Markovian dynamics due to strong
backaction from the environment have attracted much attention in practical
implementation of nanoscale solid-state quantum information processing (QIP)
both theoretically \cite{Op1,Op2,zeno,ZWM} and experimentally \cite%
{Op3,GGC1,Mad1,Hoe,Mad2}. It is widely believed that the unavoidable
interaction with Markovian environment results in decoherence effect and
loss of quantum correlation, so the quantum correlation preservation (QCP)
in quantum systems has become a critical challenge in quantum computation
technologies. Previous studies have shown that long-time quantum
entanglement protection can be realized by engineering the structured
environment such as photonic band gap (PBG) materials, which has periodic
dielectric structures exhibiting a range of frequencies with electromagnetic
wave propagation forbidden \cite{JOB}. This novel property mostly arises
from the specific structure of PBG environment \cite{John1}, which leads to
strong emitter-field correlation and formation of emitter-field bound states
(EFBS) \cite{John2} with the spontaneous emission being greatly inhibited
\cite{Lei}. Experimentally, the non-Markovian effect of the emitter in the
PBG reservoir has been directly observed in several kinds of nanoscale
cavities \cite{Mad1,Hoe,Mad2}.

\begin{figure}[tbph]
\centering\includegraphics[width=0.95\columnwidth]{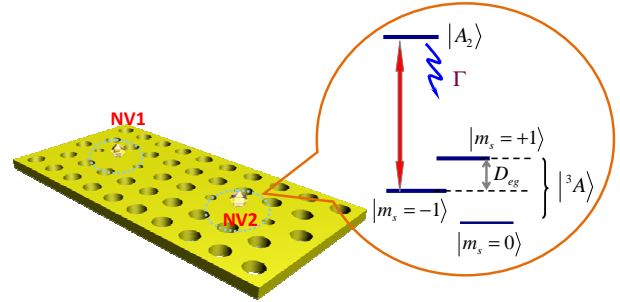}
\caption{(Color online) The composite NV-PCC system consists of a planar PCC
and two identical NVCs in diamond nanocrystals. The inset shows the level
structure of a NVC, where the electronic ground state $\left\vert
^{3}A\right\rangle$ is a spin triplet state, and $D_{eg}=\protect\gamma %
_{e}B_{0}$ is the level splitting induced by an external magnetic field $%
B_{0}$ with $\protect\gamma_{e}$ the electron gyromagnetic ratio. The red
arrow denotes the coupling between NVC and PCC. $\Gamma$ is the spontaneous
decay rate of the excited state $|A_{2}\rangle $.}
\label{scm}
\end{figure}

In this Letter, we focus on the issue of decoherence suppression in
solid-state systems consisting of diamond nitrogen-vacancy centers (NVCs)
\cite{NV1} and planar photonic crystal cavities (PCC) \cite{PC}. The NVC is
an attractive spin qubit since it exhibits a unique combination of robust
room-temperature spin coherence \cite{NV2} and efficient optical
addressability, controllability, and readout \cite{NV3,NV4,NV5,NV6}.
Furthermore, the planar PCC with high $Q$ factor (i.e., $Q>1\times 10^{6}$
even up to $2\times 10^{7}$ in the optimized structure \cite{PC1,pc1}) and
mode volumes comparable to a cubicoptical wavelength \cite{PC2} can strongly
confine photons in a tiny space of optical-wavelength dimension within a PBG
structure. Numerous experiments have successively demonstrated the strong
coupling between NVCs and the modes of silicon nitride PCC \cite{NVPC1},
gallium phosphide PCC \cite{PC1,PC2,NVPC2}, and PCC in monocrystalline
diamond \cite{NVPC3,NVPC33}, respectively. These advances imply that the
conventional Born-Markovian description to the NVC decoherence induced by
the PPC is not applicable anymore.

In the present work we investigate the non-Markovian dynamics of quantum
correlation between two initially entangled NVCs, each of which is embedded
in a nanocavity in the planar PCC coupled to the radiation fields initially
in vacuum states. Characterizing the quantum correlation by both quantum
discord (QD) \cite{QD1} and entanglement of formation (EoF) \cite{EOF}, we
find that, if the transition frequency of the NVC is fully within the band
gap of the PCC, a finite quantum correlation is preserved due to the
interplay between the non-Markovian effect of the structured reservoirs and
the existence of EFBS. Otherwise, no QCP can be observed, and there only
exists the quantum correlation between the two independent reservoirs in the
long-time limit. The essential condition for realizing this QCP is
explicitly given. These results would be useful for experimental exploration
of non-Markovian features in spin-based quantum system.

\section{The model and Hamiltonian}
Now we focus on two noninteracting NVCs (NV1 and NV2) coupled, respectively,
to two uncorrelated vacuum reservoirs $R_{1}$ and $R_{2}$, namely, two
nanocavities in a planar PCC, as shown in Fig. \ref{scm}. Each NVC is
negatively charged with two unpaired electrons located near the vacancy,
usually treated as electron spin-$1$. In our scheme, the PCC modes with $%
\sigma ^{+}$ polarization are coupled to the transition from the NVC's
ground state sublevel $\left\vert -\right\rangle =$ $\left\vert
^{3}A,m_{s}=-1\right\rangle $ to one of the excited states $\left\vert
+\right\rangle =\left\vert A_{2}\right\rangle =(\left\vert
E_{-},m_{s}=+1\right\rangle +\left\vert E_{+},m_{s}=-1\right\rangle )/\sqrt{2%
}$ \cite{NV6}. Due to the dynamical independence between the two local
subsystems, we can first solve the individual subsystem, and then apply the
result obtained to the double case. Each subsystem (i.e., a NVC plus a
radiation field propagating in the PCC) is governed by \cite{Scu}
\begin{equation}
\hat{H}=\omega _{0}\hat{\sigma}_{+}\hat{\sigma}_{-}+\sum\nolimits_{k}\omega
_{k}\hat{a}_{k}^{\dag }\hat{a}_{k}+\sum\nolimits_{k}(g_{k}\hat{\sigma}_{+}%
\hat{a}_{k}+h.c.),  \label{ham}
\end{equation}%
where $\hat{\sigma}_{\pm }$ and $\omega _{0}$ are the inversion operator and
transition frequency of the NVC, and $\hat{a}_{k}^{\dag }$ ($\hat{a}_{k})$
are the creation (annihilation) operators of the $k$-th mode of the
reservoir. The coupling strength between the NVC and the reservoir is
denoted by $g_{k}=\omega _{0}\mathbf{\vec{d}}\cdot \mathbf{\vec{e}}_{k}/%
\sqrt{2\epsilon _{0}\omega _{k}V_{0}}$ \cite{WG}, where $\mathbf{\vec{e}}%
_{k} $ and $V_{0}$ are the unit polarization vector and the normalization
volume of the radiation field, $\mathbf{\vec{d}}$ is the dipole moment of
the NVC, and $\epsilon _{0}$ is the free space permittivity. Here the
specific periodic structure of the PCC causes a band-gap dispersion relation
to the field \cite{John3}.
\begin{equation}
\omega _{k}=\omega _{c}+A(k-k_{0})^{2},  \label{disp}
\end{equation}%
where $A\simeq $ $\omega _{c}/k_{0}^{2}$ with $k_{0}\simeq \omega _{c}/c$
being a specific wave vector with respect to the point-group symmetry of the
PCC, and $\omega _{c}$ is the dip of the band frequency. Note that Eq. (\ref%
{disp}) describes well our 2D PPC structure, where the 2D square lattice is
formed by cylinders (see Fig. \ref{scm}). Here the two orthogonal directions
are equivalent and the dispersion relation is the same for both.

\begin{figure}[tbph]
\centering%
\includegraphics[width=8.4 cm,bb=113pt 530pt 499pt
769pt]{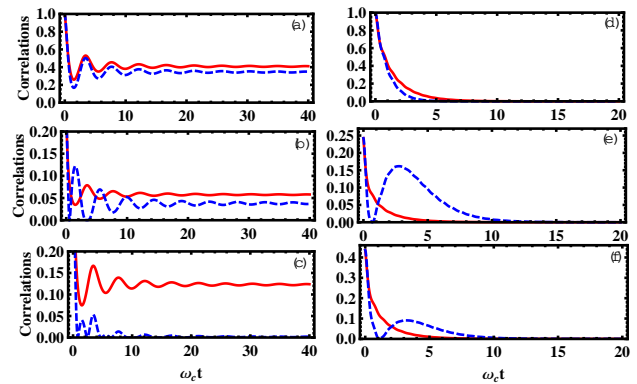}
\caption{(Color online) Time evolution of the quantum correlations between
NV1 and NV2 calculated by QD (solid-line) and EoF (dashed-line), where (a) $%
\protect\alpha =1/\protect\sqrt{2}$, and $\protect\omega _{0}=\protect\omega %
_{c}/10$; (b) $\protect\alpha =0.2$, and $\protect\omega _{0}=\protect\omega %
_{c}/10$; (c) $\protect\alpha =0.3$, and $\protect\omega _{0}=\protect\omega %
_{c}/10$; (d) $\protect\alpha =1/\protect\sqrt{2}$, and $\protect\omega %
_{0}=10\protect\omega _{c}$; (e) $\protect\alpha =0.2$, and $\protect\omega %
_{0}=10\protect\omega _{c}$; (f) $\protect\alpha =0.3$, and $\protect\omega %
_{0}=10\protect\omega _{c}$. The parameter $\protect\eta =0.2$ is used. }
\label{nvccor}
\end{figure}

\section{The dynamics of quantum correlations}
We firstly study the time evolution of a single NVC prepared initially in
the excited state $|+\rangle $ influenced by the reservoir. The state of the
system evolves as
\begin{equation}
\left\vert \psi (t)\right\rangle =b(t)\left\vert +,\{0_{k}\}\right\rangle
+\sum\nolimits_{k=0}^{\infty }b_{k}(t)\left\vert -,\{1_{k}\}\right\rangle
\text{,}  \label{amp}
\end{equation}%
where $|\{1_{k}\}\rangle $ is the single-photon state in $k$-th mode of the
reservoir. The amplitude $b(t)$ at any time satisfies following
integro-differential equation
\begin{equation}
\dot{b}(t)+i\omega _{0}b(t)+\int_{0}^{t}f(t-\tau )b(\tau )d\tau =0\text{,}
\label{itdg}
\end{equation}%
where the correlation function is $f(t-\tau )=\eta \int \frac{k^{2}c^{3}}{%
\omega _{k}}e^{-i\omega _{k}(t-\tau )}dk$. The non-Markovian memory effect
of the structured reservoir has been registered self-consistently in the
kernel function $f(t-\tau )$ in Eq. (\ref{itdg}). Going back to the
double-NVC case, the master equation of NVCs can be derived by tracing over
the environmental degrees of freedom from Eq. (\ref{amp}) \cite{Op1},
\begin{eqnarray}
\dot{\rho}(t) &=&\sum_{n=1}^{2}\{-i\Omega (t)[\sigma _{+}^{n}\sigma
_{-}^{n},\rho (t)]+\gamma (t)[2\sigma _{-}^{n}\rho (t)\sigma _{+}^{n}  \notag
\\
&&-\sigma _{+}^{n}\sigma _{-}^{n}\rho (t)-\rho (t)\sigma _{+}^{n}\sigma
_{-}^{n}]\},  \label{master}
\end{eqnarray}%
where $\Omega (t)=-\text{Im}[\frac{\dot{b}(t)}{b(t)}]$ and $\gamma (t)=-%
\text{Re}[\frac{\dot{b}(t)}{b(t)}]$ denote Lamb shifted frequency and decay
rate of the NVCs, respectively. If $f(t-\tau )$ in Eq. (\ref{itdg}) is
replaced by a \textit{delta} function, then Eq. (\ref{master}) recovers the
conventional master equation under Born-Markovian approximation \cite{BMA}.

Starting from the initial state of the whole system as $|\Psi (0)\rangle
=(\alpha |-,-\rangle +\beta |+,+\rangle )|\{0_{k}\}\rangle
_{r_{1}}|\{0_{k}\}\rangle _{r_{2}}$ \cite{explain}, we can obtain the
time-dependent state as%
\begin{equation}
|\Psi (t)\rangle =\alpha |-,\{0_{k}\}\rangle _{1}|-,\{0_{k}\}\rangle
_{2}+\beta |\psi (t)\rangle _{1}|\psi (t)\rangle _{2},  \label{tds}
\end{equation}%
where Eq. (\ref{amp}) has been rewritten as $|\psi (t)\rangle
=b(t)\left\vert +\right\rangle |\bar{0}\rangle _{r}+\tilde{b}(t)\left\vert
-\right\rangle |\bar{1}\rangle _{r}$ with the collective states of the
reservoir defined as $|\bar{0}\rangle _{r}=|\{0_{k}\}\rangle $ and $|\bar{1}%
\rangle _{r}=\frac{1}{\tilde{b}(t)}\sum\nolimits_{k}b_{k}(t)|1_{k}\rangle $
and $\tilde{b}(t)=\sqrt{1-\left\vert b(t)\right\vert ^{2}}$ \cite{Lop}. From
Eq. (\ref{tds}) the reduced density matrix $\rho _{N_{1}N_{2}}$ for the
subsystem NV1-NV2 can be obtained by tracing over the degrees of freedom of
the two reservoirs,
\begin{equation}
\rho _{N_{1}N_{2}}(t)=\left(
\begin{array}{cccc}
|\beta |^{2}\left\vert b(t)\right\vert ^{4} & 0 & 0 & \beta \alpha ^{\ast
}b(t)^{2} \\
0 & p & 0 & 0 \\
0 & 0 & p & 0 \\
\beta ^{\ast }\alpha b^{\ast }(t)^{2} & 0 & 0 & q%
\end{array}%
\right) ,
\end{equation}%
with $p=|\beta b(t)|^{2}\tilde{b}(t)^{2}$ and $q=1-|\beta |^{2}|b(t)|^{4}-2p$%
. Similarly, the corresponding reduced density matrices can be obtained for
other subsystems like reservoir1-reservoir2 ($r_{1}r_{2}$) and NVC-reservoir
($N_{1}r_{1}$, $N_{1}r_{2}$, $N_{2}r_{1}$, $N_{2}r_{2}$).

The quantum correlations can be quantified by EoF \cite{EOF} and QD \cite%
{QD1,QD2}. The former is defined as $E(\rho )=H[\frac{1+\sqrt{1-C(\rho )^{2}}%
}{2}]$ with $H[x]=-x\log _{2}x-(1-x)\log _{2}(1-x)$ and $C(\rho )=\max \{0,%
\sqrt{\lambda _{1}}-\sqrt{\lambda _{2}}-\sqrt{\lambda _{3}}-\sqrt{\lambda
_{4}}\}$, where the decreasing-order-arranged quantities $\lambda _{i}$ are
the eigenvalues of the matrix $\rho (\sigma _{y}^{A}\otimes \sigma
_{y}^{B})\rho ^{\ast }(\sigma _{y}^{A}\otimes \sigma _{y}^{B})$ with $\rho
^{\ast }$ the complex conjugation of $\rho $ and $\sigma _{y}^{A(B)}$ the
Pauli matrix acting on the subsystem $A(B)$ \cite{Woo}. On the other hand,
QD is defined as the minimum difference between two ways defining mutual
information (MI), $Q(\rho )=I(\rho _{AB})-\max_{\{\Pi _{k}\}}I(\rho
_{A}|\{\Pi _{k}\})$. Here $I(\rho _{AB})=S(\rho _{A})+S(\rho _{B})-S(\rho
_{AB})$ is the quantum MI and $\max_{\{\Pi _{k}\}}I(\rho _{A}|\{\Pi
_{k}\})=\max_{\{\Pi _{k}\}}[S(\rho _{A})-\sum\nolimits_{k}p_{k}S(\rho
_{A}^{k}|\Pi _{k})]$ is the maximal MI when a measurement is performed on
subsystem $B$ \cite{QD1}, where $S(\cdot ) $ is the von Neumann entropy, $%
\{\Pi _{k}\}$ is a completely positive-operator-valued measure on the
subsystem $B$ and $p_{k}$ is the respective probabilities.

Fig. \ref{nvccor} presents the quantum correlation dynamics (EoF and QD) of
the NV1-NV2 subsystem with different initial states, where two typical cases
of $\omega_{0}<\omega_{c}$ and $\omega_{0}\gg\omega_{c}$ are considered,
corresponding to the NVC's transition frequency within and far beyond the
band gap of the PBG material, respectively. In the former case, as shown in
Fig. \ref{nvccor}(a,b,c), the correlations calculated by EoF and QD exhibit
obvious oscillation due to the energy exchange between the NVC and the
memory reservoir, and then the correlations approach a definite value in the
long-time limit. It implies that the decay rate of the excited state $%
\left\vert A_{2}\right\rangle $ of the NVC approaches zero after some
oscillations, which leads to the QCP. However, in the $\omega _{0}\gg\omega
_{c}$ case, the residual correlation by EoF and QD between NV1 and NV2
vanishes in the long-time limit, although there exists transient revival in
EoF [see Fig. \ref{nvccor}(e,f)]. All the\ differences between these two
cases indicate that the PBG plays a vital role in the quantum correlation
evolution of the NVCs confined in the structured environment. We will return
to this point later.

\begin{figure}[tbph]
\centering\includegraphics[width=8 cm,bb=88pt 250pt 538pt
761pt]{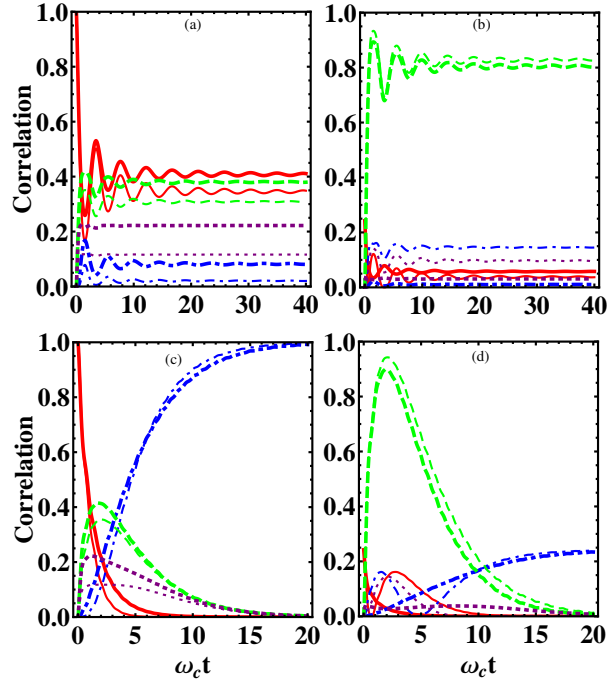}
\caption{(Color online) Time evolution of the quantum correlations
calculated by QD (thick-line) and EoF (thin-line) for different subsystems,
where the thick-solid line, thick-dot-dashed line, thick-dashed line, and
thick-dotted line denote the QD1, QD2, QD3, and QD4, respectively. The
thin-solid line, thin-dot-dashed line, thin-dashed line, and thin-dotted
line denote the Eof1, Eof2, Eof3, and Eof4, respectively. (a) $\alpha =1/\protect\sqrt{2}$, and $\omega _{0}=\omega _{c}/10$%
; (b) $\alpha =0.2$, and $\omega _{0}=\omega _{c}/10$%
; (c) $\alpha =1/\sqrt{2}$, and $\omega _{0}=10%
\omega _{c}$; (d) $\alpha =0.2$, and $\omega _{0}=10%
\omega _{c}$. The parameter $\eta =0.2$ is used. }
\label{compcor}
\end{figure}

To get a clear picture on how the quantum correlation is distributed among
different subsystems, we have calculated quantum correlations for different
bipartite partitions, such as reservoir1-reservoir2, NV1-reservoir1, and
NV1-reservoir2 labeled by QD2 (Eof2), QD3 (EoF3) and QD4 (EoF4),
respectively, and the quantum correlations between NV1-NV2 are denoted by
QD1 (EoF1). Fig. \ref{compcor} shows the time evolution of EoF and QD in
each subsystem under different initial states, and one can find that for
each partition, the correlations denoted by EoF and QD behave similarly with
obvious oscillations. Interestingly, once the two NVCs are initially
prepared in maximally entangled state with $\alpha=\beta=1/\sqrt{2}$, QD is
always larger than EoF for both $\omega_{0}<\omega_{c}$ and $%
\omega_{0}\gg\omega_{c}$ cases, as shown in Fig. \ref{compcor}(a,c). The
only exception is the curves for QD2 and EoF2 in Fig. \ref{compcor}(c) in
the $\omega_{0}\gg\omega_{c}$ case, which do not obey the monogamic
relation. In contrast to this feature, if the NVCs are not initially
prepared in maximally entangled states, as in Fig. \ref{compcor}(b,d), QD is
usually smaller than EoF. Furthermore, it is important to note that the QCP
in each subsystem exists only for the NV's transition frequency within the
band gap, as shown in Fig. \ref{compcor}(a,b). Another obvious feature is
that the quantum correlation initially prepared between NVCs distributes
between other subsystems with different weights. However, for $%
\omega_{0}\gg\omega_{c}$, the residual quantum correlation of QD1, QD3, and
QD4 vanishes in the long-time limit, but the quantum correlations (QD2 and
EoF2) between the two independent reservoirs approach their maximum values,
which implies that quantum correlation initially prepared in NVCs has been
fully transferred to the reservoirs, significantly different from the
multi-site distribution of quantum correlation in the case of $%
\omega_{0}<\omega_{c}$.

\section{The discussion and conclusion}
The Born-Markovian approximation is only valid in our case when the
correlation function of the radiation field has an infinitesimal time scale,
which is always valid when its density of states is a continuous and
smoothly varying function in the frequency space \cite{Veg}. However, the
distribution function for PBG reservoir is not a monotonous function due to
the existence of band edge, where the density of states varies rapidly with
a rate comparable to the NVC's spontaneous emission rate \cite{John2}, then
it becomes non-exponential and the emission spectrum becomes non-Lorentzian.
By accurately calculating the non-Markovian dynamics, we really find that
the quantum correlation dynamics displays an oscillatory behavior and
non-exponential tendency. Figs. \ref{nvccor} and \ref{compcor} indicate
clearly that the photon emitted through spontaneous emission goes back and
forth between the NVCs and their local reservoirs. It means that the
environment is not robust any more during the interaction with the quantum
system, but changes at the time scale of its memory time to reach a new
equilibrium state.

\begin{figure}[tbp]
\centering
\includegraphics[width=0.47\columnwidth]{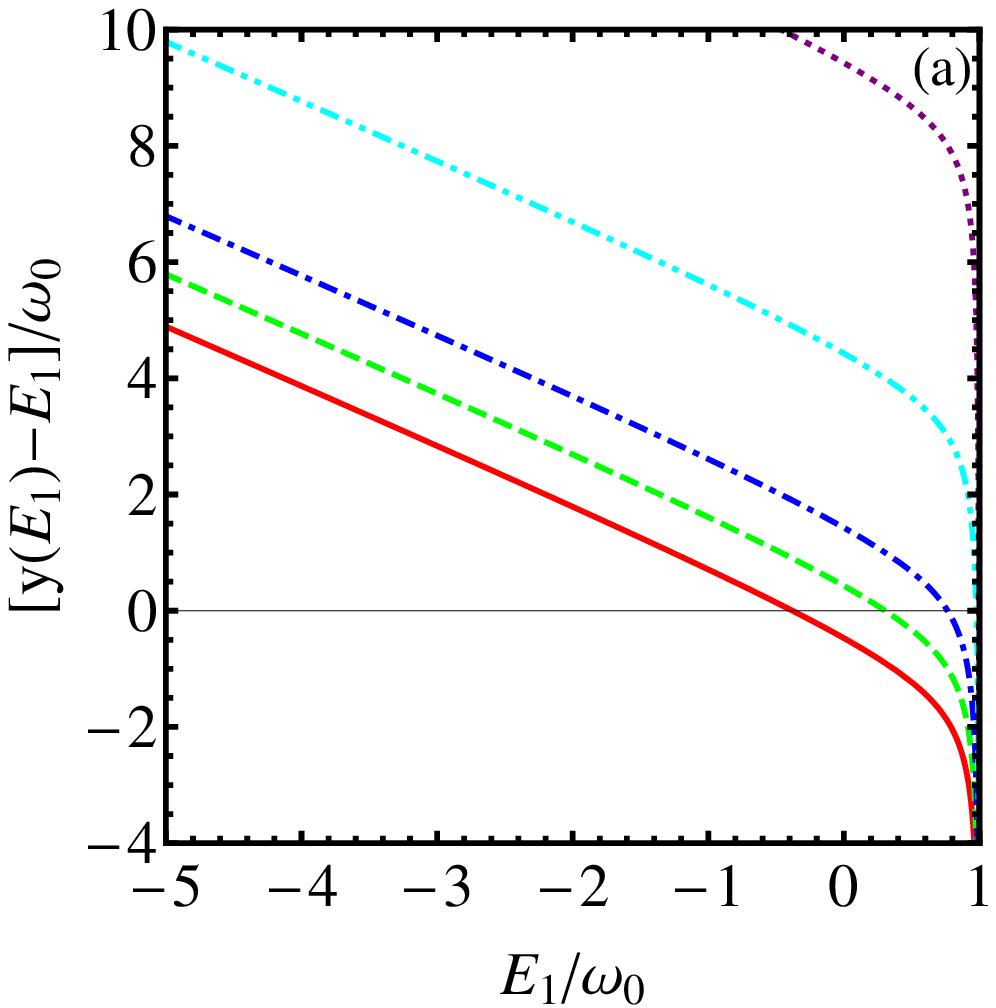}~~~%
\includegraphics[width=0.47\columnwidth]{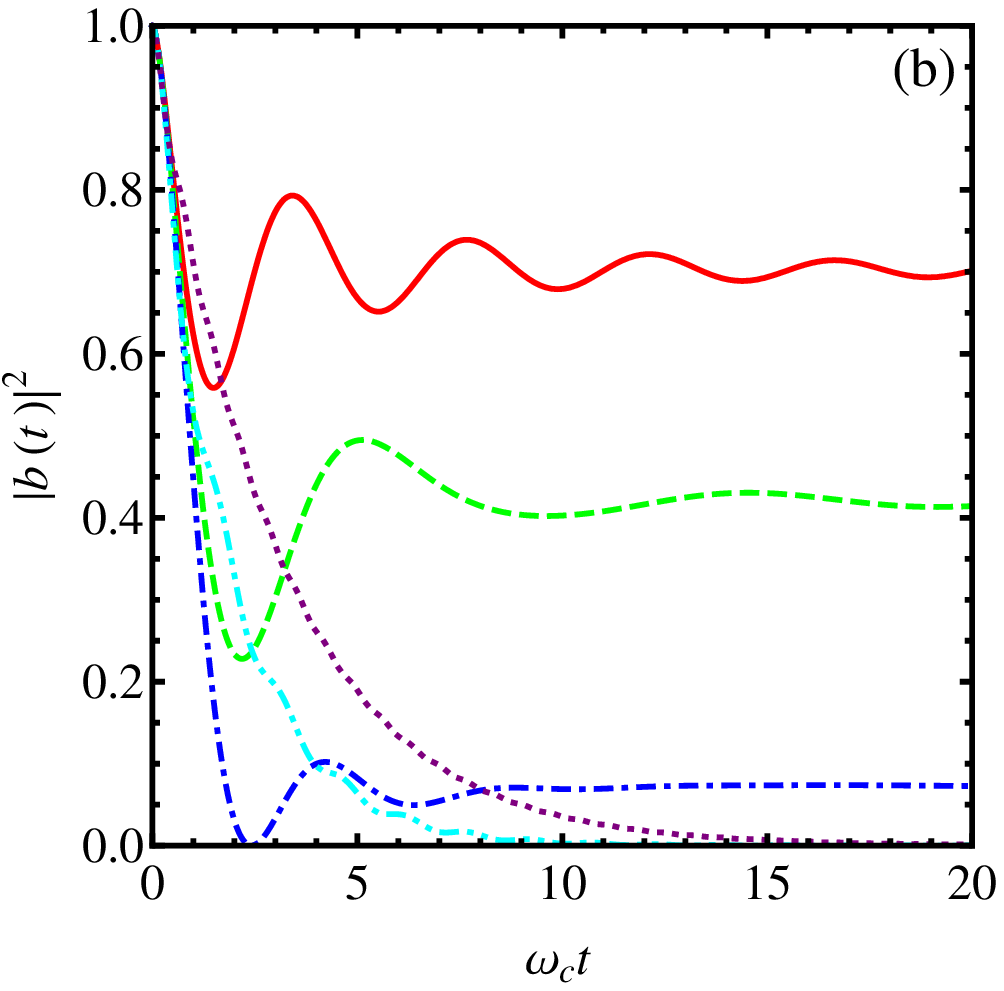}
\caption{(Color online) Illustration of the formation of EFBS (a) and its
dynamical consequence to the excited-state population (b) when $\protect%
\omega _{0}=\protect\omega _{c}/10$ (solid line), $\protect\omega _{c}$
(dashed line), $2\protect\omega _{c}$ (dot-dashed line), $5\protect\omega %
_{c}$ (dot-dot-dashed line), and $10.0\protect\omega _{c}$ (dotted line).
The intersect of $y(E_{1})-E_{1}$ with the horizontal axis characterizes the
formation of EFBS. $\protect\eta =0.2$ is used. }
\label{bds}
\end{figure}

We attribute the mechanism of this novel QCP to the interplay between the
formed EFBS and the non-Markovian effect. Here we argue that there are two
essential conditions for realizing the QCP. One is the existence of the
EFBS, which provides the ability to preserve quantum correlation, and the
other is the non-Markovian effect, which provides a dynamical way to
preserve the quantum correlation \cite{An}. Thus, whether the EFBS exists or
not plays a key role in achieving the QCP, even if the reservoir has already
shown the memory effect.

To verify this argument, we find under what condition the EFBS can be
formed. The total excitation number $\hat{N}=\hat{\sigma}_{+}\hat{\sigma}%
_{-}+\sum_{k}\hat{a}_{k}^{\dag }\hat{a}_{k}$ of Eq. (\ref{ham}) is
conserved, so the Hilbert space is split into the direct sum of the
subspaces with definite $N$. From the eigen equation in the $N=1$ subspace,
we have the eigenvalue $E_{1}$ fulfilling
\begin{equation}
y(E_{1})\equiv \omega _{0}+\eta \int_{0}^{\infty }{\frac{c^{3}k^{2}}{%
(E_{1}-\omega _{k})\omega _{k}}}dk=E_{1}\text{,}  \label{eigen}
\end{equation}%
where $\eta =\omega _{0}^{2}d^{2}/6\pi ^{2}\epsilon _{0}c^{3}$ is a
dimensionless coupling constant, and $\eta \omega _{0}$ is the natural
spontaneous emission rate of the excited state $|A_{2}\rangle $ of NVC.
Combining with Eq. (\ref{disp}), we can find that $y(E_{1})$ is
monotonically decreasing in the region $E_{1}\in (-\infty ,\omega _{c})$,
i.e. in the band gap of the reservoir. It means that Eq. (\ref{eigen}) may
have one and only one solution in this region if the system parameters
fulfill $y(\omega _{c})<\omega _{c}$ [see Fig. \ref{bds}(a)]. The curves
when $\omega _{0}$ is smaller than or comparable with $\omega _{c}$
definitively have an intersection point with the horizontal line so that one
discrete root exists for Eq. (\ref{eigen}). We call this discrete eigenstate
as the EFBS. On the other hand, no root exists in the region $(\omega
_{c},+\infty )$ because it would make $y(E_{1})$ divergent otherwise. As a
stationary state of the whole system, the EFBS does not lose any quantum
coherence during the time evolution, implying that the decoherence can be
inhibited. Fig. \ref{bds}(b) shows the dynamical consequence of the EFBS on
the excited-state population of the NVC. We really find that once the EFBS
is formed, the excited-state population will stabilize to a steady value in
the long-time limit.

The underlying physical picture is that during the evolution the NVC's
excited level $\left\vert A_{2}\right\rangle $ experiences an anomalous
``giant" Lamb shift and splits into doublet levels only when the NVC's
transition frequency $\omega _{0}$ lies inside or near the band gap \cite%
{John4,Mog,FQ}. One of the doublet levels retains in EFBS lying within the
gap, whereas the other is shifted out of the gap and exhibits resonance
fluorescence. In the $\omega _{0}<\omega _{c}$ case, once the FEBS has been
established, the excited-state population of the NVC in a FEBS remains
constant in time because the FEBS state is actually a stationary state with
a vanishing decay rate during the time evolution. The claim above can be
further verified by the fact that the QCP is always absent in the $%
\omega_{0}\gg\omega_{c}$ case, as shown in the right panels of Fig. \ref%
{nvccor} and in the lower panels of Fig. \ref{compcor}, where the EFBS does
not appear due to the NVC's transition frequency far beyond the upper band
of the PBG. It is physically originated from the fact that the eigenstate of
the Hamiltonian (\ref{compcor}) with the eigenvalues resides in a continuous
band experiences an out-of-phase interference during the time evolution,
which makes the excited-state population approaching zero asymptotically and
causes the severe decoherence to the NV1-NV2 subsystem.

In summary, we have investigated the dynamics of quantum correlation between
two initially entangled NVCs embedded, respectively, in PCC. We have shown
the existence of QCP resulted from both the memory effect of the
non-Markovian structured reservoir and the EFBS. With the mechanism of QCP,
we expect to have a significant step toward the future full-scale quantum
information processor based on the increasingly-developed nanoscale
solid-state technology. It is noted that although only 2D structure of
photonic crystal is taken into accounted in our work, our result could
potentially be applicable to the 3D case with a more complicated PBG
structure \cite{Hoe}. We expect that the QCP could be achieved whenever the
EFBS is formed under the condition that the transition frequency of the NVC
is within the band gap of the 3D photonic crystal. The qualitative
verification to this expectation is worth to be performed by combing the
explicit form of 3D photonic crystal structure in the future.

\section*{Acknowledgements}
The authors thank Z. Y. Xu, Q. J. Tong, and Z. Q. Yin for enlightening
discussions. This work is supported partially by the National Fundamental
Research Program of China under Grant No. 2012CB922102, by the NNSF of China
under Grants No. 11274351, No. 10974225, No. 11004226, No. 11104326, No.
11105136, No. 11175072 and No. 11204196, by the Specialized Research Fund
for the Doctoral Program of Higher Education, and by the National Research
Foundation and Ministry of Education, Singapore (Grant No. WBS:
R-710-000-008-271).


\begin{thebibliography}{99}
\bibitem{Op1} H.-P Breuer and F. Petruccione, \textit{The Theory of Open
Quantum Systems} (Oxford University, Oxford, 2007).

\bibitem{Op2} P. Kaer, T. R. Nielsen, P. Lodahl, A.-P. Jauho, and J. M\o rk,
Phys. Rev. Lett. \textbf{104}, 157401 (2010); M. \v{Z}nidari\v{c}, C.
Pineda, and I. Garc\'{\i}a-Mata, Phys. Rev. Lett. \textbf{107}, 080404
(2011); E.-M. Laine, H.-P. Breuer, J. Piilo, C.-F. Li, and G.-C. Guo, Phys.
Rev. Lett. \textbf{108}, 210402 (2012); A. Carmele, A. Knorr, and F. Milde,
arXiv:1203.0126.

\bibitem{zeno} S. Maniscalco, F. Francica, R. L. Zaffino, N. L. Gullo, and
F. Plastina, Phys. Rev. Lett. \textbf{100}, 090503 (2008); D. Alonso and In%
\'{e}s de Vega, Phys. Rev. Lett. \textbf{94}, 200403 (2005); J. Piilo, S.
Maniscalco, K. H\"{a}rk\"{o}nen, and K.-A. Suominen, Phys. Rev. Lett.
\textbf{100}, 180402 (2008).

\bibitem{ZWM} W.-M. Zhang, P.-Y. Lo, H.-N. Xiong, M. W.-Y. Tu, and F. Nori,
Phys. Rev. Lett. \textbf{109}, 170402 (2012); L. Ferialdi and A. Bassi,
Phys. Rev. Lett. \textbf{108}, 170404 (2012); S. F. Huelga, \'{A}.
Rivas, and M. B. Plenio, Phys. Rev. Lett. \textbf{108}, 160402 (2012).

\bibitem{Op3} H. Tahara, Y. Ogawa, and F. Minami, Phys. Rev. Lett. \textbf{%
107}, 037402 (2011); C. Galland, A. H\"{o}gele, H. E. T\"{u}reci, and A.
Imamo\u{g}lu, Phys. Rev. Lett. \textbf{101}, 067402 (2008).

\bibitem{GGC1} B.-H. Liu, L. Li, Y.-F. Huang, C.-F. Li, G.-C. Guo, E.-M.
Laine, H.-P. Breuer, and J. Piilo, Nat. Phys. \textbf{7}, 931 (2011); B.-H.
Liu, D.-Y. Cao, Y.-F. Huang, C.-F. Li, G.-C. Guo, E.-M. Laine, H.-P. Breuer,
J. Piilo, arXiv:1208.1358.

\bibitem{Mad1} K. H. Madsen, S. Ates, T. Lund-Hansen, A. L\"{o}ffler, S.
Reitzenstein, A. Forchel, and P. Lodahl, Phys. Rev. Lett. \textbf{106},
233601 (2011).

\bibitem{Hoe} U. Hoeppe, C. Wolff, J. K\"{u}chenmeister, J. Niegemann, M.
Drescher, H. Benner, and K. Busch, Phys. Rev. Lett. \textbf{108}, 043603
(2012).

\bibitem{Mad2} K. H. Madsen, P. Kaer, A. Kreiner-M\o ller, S. Stobbe, A.
Nysteen, J. M\o rk, P. Lodahl, arXiv:1205.5623.

\bibitem{JOB} M. Fujita, S. Takahashi, Y. Tanaka, T. Asano, and S. Noda,
Science \textbf{308}, 1296 (2005).

\bibitem{John1} N. Vats and S. John, Phys. Rev. A \textbf{58}, 4168 (1998).

\bibitem{John2} S. John, Phys. Rev. Lett. \textbf{58}, 2486 (1987); S. John
and J. Wang, \textit{ibid}. \textbf{64}, 2418 (1990); S. John and T. Quang,
\textit{ibid}. \textbf{74}, 3419 (1995).

\bibitem{Lei} M. D. Leistikow, A. P. Mosk, E. Yeganegi, S. R. Huisman, A.
Lagendijk, and W. L. Vos, Phys. Rev. Lett. \textbf{107}, 193903 (2011); M.
R. Jorgensen, J. W. Galusha, and M. H. Bartl, Phys. Rev. Lett. \textbf{107},
143902 (2011); H. Ning and P. V. Braun, Physics \textbf{4}, 76 (2011).

\bibitem{NV1} F. Jelezko, T. Gaebel, I. Popa, A. Gruber, and J. Wrachtrup,
Phys. Rev. Lett. \textbf{92}, 076401 (2004); D. M. Toyli, D. J. Christle, A.
Alkauskas, B. B. Buckley, C. G. Van de Walle, and D. D. Awschalom, Phys.
Rev. X \textbf{2}, 031001 (2012).

\bibitem{PC} B.-S. Song, S. Noda, T. Asano, and Y. Akahane, Nat. Mater.
\textbf{4}, 207 (2005); J. H. Hartmann, F. G. S. L. Brand\"{a}o, and M. B.
Plenio, Nat. Phys. \textbf{2}, 849 (2006); D. van Oosten and L. Kuipers,
Phys. Rev. A \textbf{84}, 011802(R) (2011).

\bibitem{NV2} P. C. Maurer, G. Kucsko, C. Latta, L. Jiang, N. Y. Yao, S. D.
Bennett, F. Pastawski, D. Hunger, N. Chisholm, M. Markham, D. J. Twitchen,
J. I. Cirac, and M. D. Lukin, Science \textbf{336}, 1283 (2012); G.
Balasubramanian, P. Neumann, D. Twitchen, M. Markham, R. Kolesov, N.
Mizuochi, J. Isoya, J. Achard, J. Beck, J. Tissler, V. Jacques, P. R.
Hemmer, F. Jelezko, and J. Wrachtrup, Nat. Mater. \textbf{8}, 383 (2009).

\bibitem{NV3} P. Neumann, J. Beck, M. Steiner, F. Rempp, H. Fedder, P. R.
Hemmer, J. Wrachtrup, and F. Jelezko, Science \textbf{329}, 542 (2010);
B.\thinspace B. Buckley, G. D. Fuchs, L. C. Bassett, and D. D. Awschalom,
Science \textbf{330}, 1212 (2010); L. Robledo, L. Childress, H. Bernien, B.
Hensen, P. F. A. Alkemade, and R. Hanson, Nature (London) \textbf{477}, 574
(2011); A. Gruber, A. Dr\"{a}benstedt, C. Tietz, L. Fleury, J. Wrachtrup,
and C. von Borczyskowski, Science \textbf{276}, 2012 (1997).

\bibitem{NV4} T. van der Sar, Z. H. Wang, M. S. Blok, H. Bernien, T. H.
Taminiau, D. M. Toyli, D. A. Lidar, D. D. Awschalom, R. Hanson, and V. V.
Dobrovitski, Nature (London) \textbf{484}, 82 (2012); P. Huang, X. Kong, N.
Zhao, F. Shi, P. Wang, X. Rong, R.-B. Liu, and J. Du, Nat. Commun. \textbf{2}%
, 570 (2011); M.\thinspace V. Dutt, L. Childress, L. Jiang, E. Togan, J.
Maze, F. Jelezko, A. S. Zibrov, P. R. Hemmer, and M. D. Lukin, Science
\textbf{316}, 1312 (2007).

\bibitem{NV5} J.\thinspace R. Maze, P. L. Stanwix, J. S. Hodges, S. Hong, J.
M. Taylor, P. Cappellaro, L. Jiang, M. V. Gurudev Dutt, E. Togan, A. S.
Zibrov, A. Yacoby, R. L. Walsworth, and M. D. Lukin, Nature (London) \textbf{%
455}, 644 (2008); E. Togan, Y. Chu, A. Imamo\u{g}lu, and M. D. Lukin, Nature
(London) \textbf{478}, 497 (2011); F. Dolde, H. Fedder, M. W. Doherty, T. N\"{o}bauer, F. Rempp, G. Balasubramanian, T. Wolf, F. Reinhard, L. C. L.
Hollenberg, F. Jelezko, and J. Wrachtrup, Nat. Phys. \textbf{7}, 459 (2011);
G. Goldstein, P. Cappellaro, J. R. Maze, J. S. Hodges, L. Jiang, A. S. S\o %
rensen, and M. D. Lukin, Phys. Rev. Lett. \textbf{106}, 140502 (2011).

\bibitem{NV6} E. Togan, Y. Chu, A. S. Trifonov, L. Jiang, J. R. Maze, L.
Childress, M. V. G. Dutt, A. S. S\o rensen, P. R. Hemmer, A. S. Zibrov, and
M. D. Lukin,\ Nature (London) \textbf{466}, 730 (2010).

\bibitem{PC1} J. Wolters, A. W. Schell, G. Kewes, N. N\"{u}sse, M.
Schoengen, H. D\"{o}scher, T. Hannappel, B. L\"{o}chel, M. Barth, and O.
Benson, Appl. Phys. Lett. \textbf{97}, 141108 (2010).

\bibitem{pc1} T. van der Sar, J. Hagemeier, W. Pfaff, E. C. Heeres, S. M.
Thon, H. Kim, P. M. Petroff, T. H. Oosterkamp, D. Bouwmeester, and R.
Hanson, Appl. Phys. Lett. \textbf{98}, 193103 (2011).

\bibitem{PC2} D. Englund, B. Shields, K. Rivoire, F. Hatami, J. Vu\v{c}kovi%
\'{c}, H. Park, and M. D. Lukin, Nano Lett. \textbf{10}, 3922 (2010).

\bibitem{NVPC1} M. W. McCutcheon and M. Lon\v{c}ar, Opt. Express \textbf{16}%
, 19136 (2008).

\bibitem{NVPC2} P. E. Barclay, K. M. Fu, C. Santori, and R. G. Beausoleil,
Opt. Express \textbf{17}, 9588 (2009).

\bibitem{NVPC3} A. Faraon, C. Santori, Z. Huang, V. M. Acosta, and R. G.
Beausolei, Phys. Rev. Lett. \textbf{109}, 033604 (2012).

\bibitem{NVPC33} J. Wolters, G. Kewes, A. W. Schell, N. N\"{u}sse, M.
Schoengen, B. L\"{o}chel, T. Hanke, R. Bratschitsch, A. Leitenstorfer, T.
Aichele, and O. Benson, Phys. Status Solidi B \textbf{249}, 918 (2012).

\bibitem{QD1} H. Ollivier and W. Zurek, Phys. Rev. Lett. \textbf{88}, 017901
(2001); L. Henderson and V. Vedral, J. Phys. A: Math. Gen. \textbf{34}, 6899
(2001).

\bibitem{EOF} C. H. Bennett, D. P. DiVincenzo, J. A. Smolin, and W. K.
Wootters, Phys. Rev. A \textbf{54}, 3824 (1996).

\bibitem{Scu} M. O. Scully and M. S. Zubairy, \textit{Quantum Optics}
(Cambridge University Press, Cambridge, 1997).

\bibitem{WG} S. M. Spillane, T. J. Kippenberg, K. J. Vahala, K. W. Goh, E.
Wilcut, and H. J. Kimble, Phys. Rev. A \textbf{71}, 013817 (2005).

\bibitem{John3} S. John and T. Quang, Phys. Rev. A \textbf{50}, 1764 (1994).

\bibitem{BMA} G. S. Agarwal, \textit{Quantum Optics} (Springer-Verlag,
Berlin, 1974); K. M\o lmer and S. Bay, Phys. Rev. A \textbf{59}, 904 (1999).

\bibitem{explain} The preparation of initial entangled state could be
completed by making the two seperated NVCs sharing a common
single-excitation, which is initially prepared in a NVC. The
single-excitation will ultimately tunnel into two emitters with a
small/large speed in the presence of the weak/strong cavity-cavity tunneling
effect during the evolution process. The relevant details can be found in
the paper \cite{YWL}.

\bibitem{YWL} W. L. Yang, Z. Q. Yin, Z. Y. Xu, M. Feng, and C. H. Oh, Phys.
Rev. A \textbf{84}, 043849 (2011).

\bibitem{Lop} C. E. L\'{o}pez, G. Romero, F. Lastra, E. Solano, and J. C.
Retamal, Phys. Rev. Lett. \textbf{101}, 080503 (2008).

\bibitem{QD2} M. F. Cornelio, M. C. de Oliveira, and F. F. Fanchini, Phys.
Rev. Lett. \textbf{107}, 020502 (2011).

\bibitem{Woo} W. K. Wootters, Phys. Rev. Lett. \textbf{80}, 2245 (1998).

\bibitem{Veg} In\'{e}s de Vega and D. Alonso, Phys. Rev. A \textbf{77},
043836 (2008).

\bibitem{An} Q.-J. Tong, J.-H. An, H.-G. Luo, and C. H. Oh, Phys. Rev. A
\textbf{81}, 052330 (2010).

\bibitem{John4} M. Woldeyohannes and S. John, J. Opt. B: Quantum Semiclass.
Opt. \textbf{5}, R43 (2003).

\bibitem{Mog} D. Mogilevtsev, F. Moreira, S. B. Cavalcanti, and S. Kilin,
Phys. Rev. A \textbf{75}, 043802 (2007).

\bibitem{FQ} B. Bellomo, R. Lo Franco, S. Maniscalco, and G. Compagno, Phys.
Rev. A \textbf{78}, 060302(R) (2008); M. Al-Amri, Gao-xiang Li, R. Tan, and
M. Suhail Zubairy, Phys. Rev. A \textbf{80}, 022314 (2009).
\end{thebibliography}
\end{document}